\documentclass[10pt,a4paper]{article}

\usepackage{amsmath}
\usepackage{amsthm}
\usepackage{mathabx}
\usepackage{url}
\numberwithin{equation}{section}
\usepackage{graphicx,lscape}
\title{\Large \bf Review  times  in  peer  review:  quantitative  analysis of editorial workflows}
\author{\small Maciej J. Mrowinski$^{1,\#}$, Agata Fronczak$^{1}$,  Piotr Fronczak$^{1}$, Olgica Nedic$^{2}$, Marcel Ausloos$^{3,4,5}$ }

 \date{\footnotesize 
 $^1$  Faculty of Physics, Warsaw University of Technology,  \\Koszykowa 75, PL-00-662, Warsaw, Poland \\($\#$) $e$-$mail$ $address$: mrow@if.pw.edu.pl \\\vskip0.2cm
 $^2$  Institute for the Application of Nuclear Energy (INEP), University of Belgrade, \\ Banatska 31b, Belgrad-Zemun, Serbia \\\vskip0.2cm
 $^3$  School of Management, University of Leicester\\ University Road, Leicester  LE1 7RH, UK;\\$e$-$mail$ $address$: ma683@le.ac.uk \\\vskip0.2cm
 $^4$  eHumanities group\footnote{Associate Researcher}$\;$, Royal Netherlands Academy of Arts and Sciences (NKVA), \\  Joan Muyskenweg 25, 1096 CJ Amsterdam, The Netherlands \\ \vskip0.2cm
 $^5$  GRAPES\footnote{Group of Researchers for Applications of Physics in Economy and Sociology}$\;$, \\ rue de la Belle Jardiniere 483, B-4031, Angleur, Belgium } 
\begin{document}
 \maketitle

\begin{abstract}
We examine selected aspects of peer review and suggest possible improvements. To this end, we analyse a dataset containing information about 300 papers submitted to the Biochemistry and Biotechnology section of the Journal of the Serbian Chemical Society. After separating the peer review process into stages that each review has to go through, we use a weighted directed graph to describe it in a probabilistic manner and test the impact of some modifications of the editorial policy on the efficiency of the whole process.

\end{abstract}
\textit{Keywords:} peer review,  editorial process, weighted directed graph
     \vskip0.5cm
    
\newpage
\section{ Introduction}\label{Introduction}
Despite a variety of criticisms of its effectiveness \cite{Wager_2001,Cooper_2009}, peer review is a fundamental mechanism for validating the quality of the research that is published in today's scientific literature \cite{Baker_2002,PRC_2008}. It is a complex, multi-phase process and there appear to be some growing concerns regarding how to improve its functioning. Given the growth of scientific journals, the increasing number of submitted articles, and the limited pool of reviewers, acquiring a good and timely review is becoming progressively more challenging. Reviews can take even a year, depending on the complexity of the topic, the number of reviewers involved, and the details of the editorial procedures.

In face of these problems, many suggestions have been proposed to make the peer review and editorial process more efficient and equitable \cite{Bornmann_2011}. In particular, the role of editors in the process of selecting and managing reviewers has been increasingly discussed \cite{Schwartz_2009,Kravitz_2010,Newton_2010}. The main focus of these discussions are ethical issues and general, qualitative recommendations for both the editors and the reviewers \cite{Cawley_2011,Resnik_2008,COPE,Wager_2006}. While such issues are certainly practical and significant, there is still the lack of quantitative suggestions that could point out possible measurable improvements to the peer review process. Do the editors send out a sufficient number of reviewer invitations to obtain two or three timely reviews of a manuscript? How often should they draw on expertise of the same reviewers consuming their time and energy? How long should they wait for a review before they can repeat an invitation or assume that a response is unlikely? What is the statistical chance that reviewers will respond? Does it depend on whether they were previously reviewers for the same journal? Although all editors try to answer these and other questions while optimizing their work on their own, they do it somewhat in the dark. Without an intensive discussion that could help to answer the aforementioned questions in a more systematic way one can be sure that editorial lags will be increasing in the years to come.

Our paper is meant to fill this gap with the help of quantitative analysis. We examine selected aspects of peer review and suggest possible improvements. To this end, we analyse a dataset containing information about 300 papers submitted to the Biochemistry and Biotechnology section of the Journal of the Serbian Chemical Society (JCSC). After separating the peer review process into stages that each review has to go through, we use a weighted directed graph to describe it in a probabilistic manner and test the impact of some modifications of the editorial policy on the efficiency of the whole process. 

The paper is organized as follows: 
  
Section 2 describes the dataset used in the paper as well as the methodology employed to analyse the data. Section 3 is devoted to the data driven theoretical analysis of the review times.  Simulations of various editorial policy scenarios and their impact on the efficiency of the process are presented in section 4. In section 5 we provide some concluding remarks and describe open problems that may be researched within the presented methodology in the future. 

\section{Review process and initial data analysis}\label{sample}

The sample we studied contains information about reviews of 58 manuscripts submitted to one of the editors of JCSC between November 2011 and July 2014. Each of 323 members of the sample corresponds to a single reviewer and comprises the group the reviewer belongs to, the ID of the reviewed manuscript and dates associated with phases of the review process. Reviewers were divided into two groups - 65 \textbf{trusted} reviewers are known personally by the editor while 258 \textbf{other} reviewers were chosen through various different means.

The review process itself is separable into distinct phases that mirror interactions between the editor, authors and reviewers. It begins when the editor, after receiving a new submission, sends out invitations to a number of reviewers (5 on average - 4 \textbf{other} and 1 \textbf{trusted}) and waits for their responses. If any of the invited reviewers does not respond, then after about 7 days an inquiry is sent. If that inquiry also remains without an answer for 10 days, then the review process for that particular reviewer is considered finished with a negative outcome. After receiving the initial invitation or the inquiry, reviewers who do answer either confirm their willingness to write the review or decline. In the latter case, much like for reviewers who did not answer at all, the review process is considered finished with a negative outcome. In the former, the editor waits for the report for 25 days before sending an inquiry. This may result in either the reviewer finishing the review and sending the report - which is the only outcome of the process that is considered positive - or a lack of answer. To sum it up, there are three possible outcomes of the review process - \textbf{report}, \textbf{no response} or \textbf{decline}.

\begin{figure}
	\centering
	\includegraphics[width=8cm]{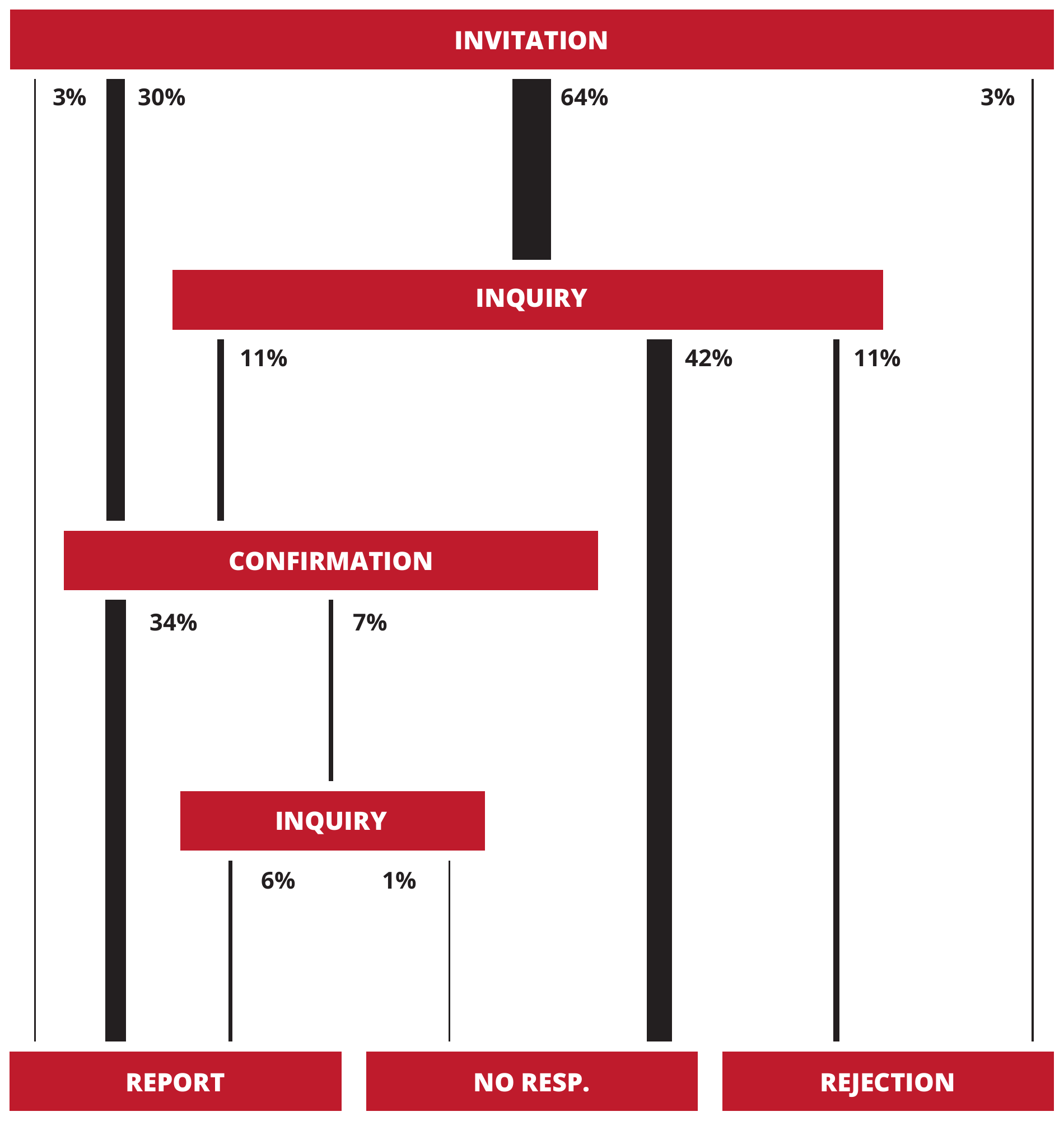}
	\caption{A graph corresponding to the review process with \textbf{trusted} and \textbf{other} reviewers. Next to each edge are probabilities of a realisation of this process passing through the edge.}
	\label{fig:bdiag-all}
\end{figure}
\begin{figure}
	\centering
	\includegraphics[width=8cm]{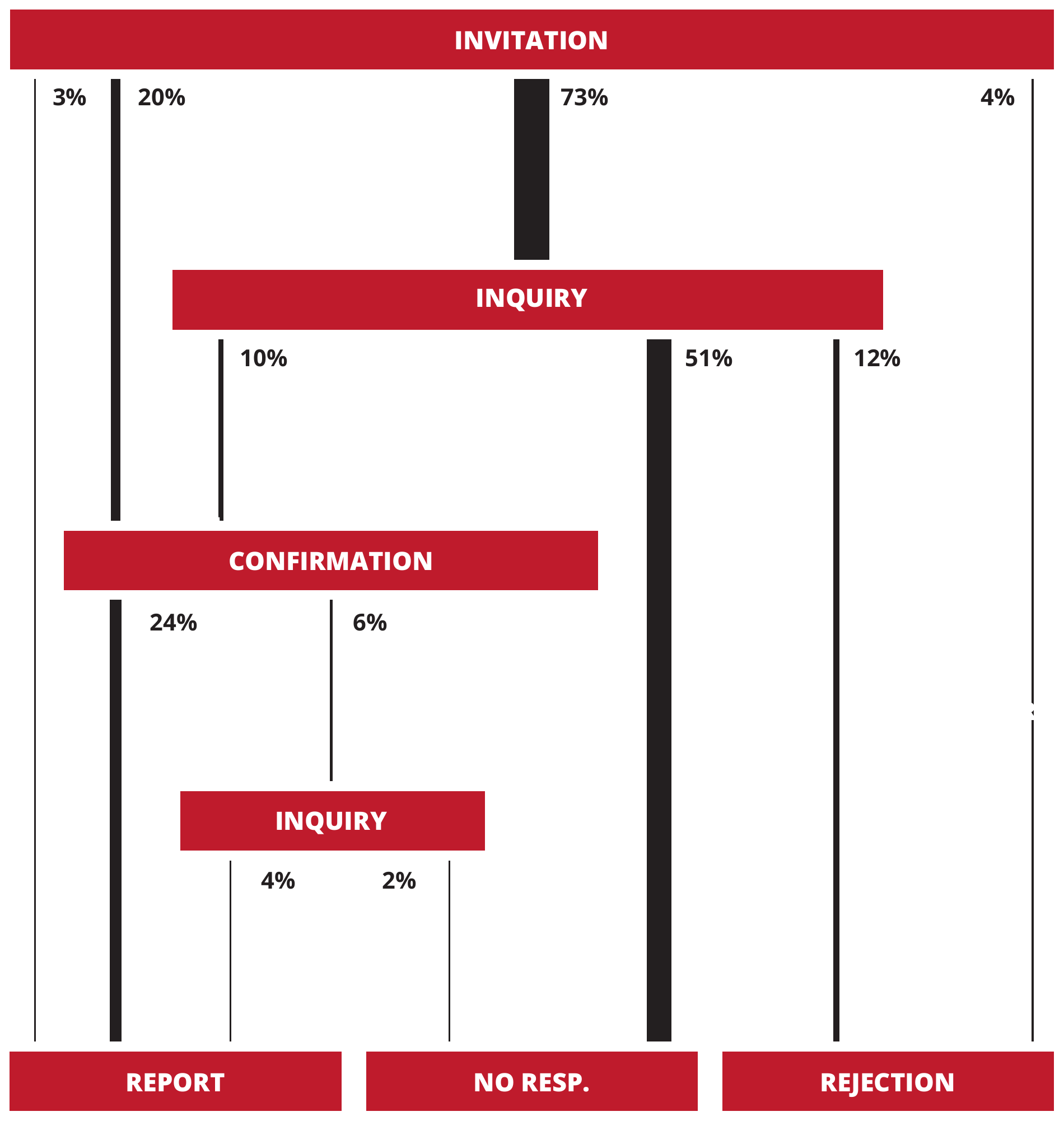}
	\caption{A graph corresponding to the review process with only \textbf{other} reviewers. Next to each edge are probabilities of a realisation of this process passing through the edge.}
	\label{fig:bdiag-other}
\end{figure}
\begin{figure}
	\centering
	\includegraphics[width=8cm]{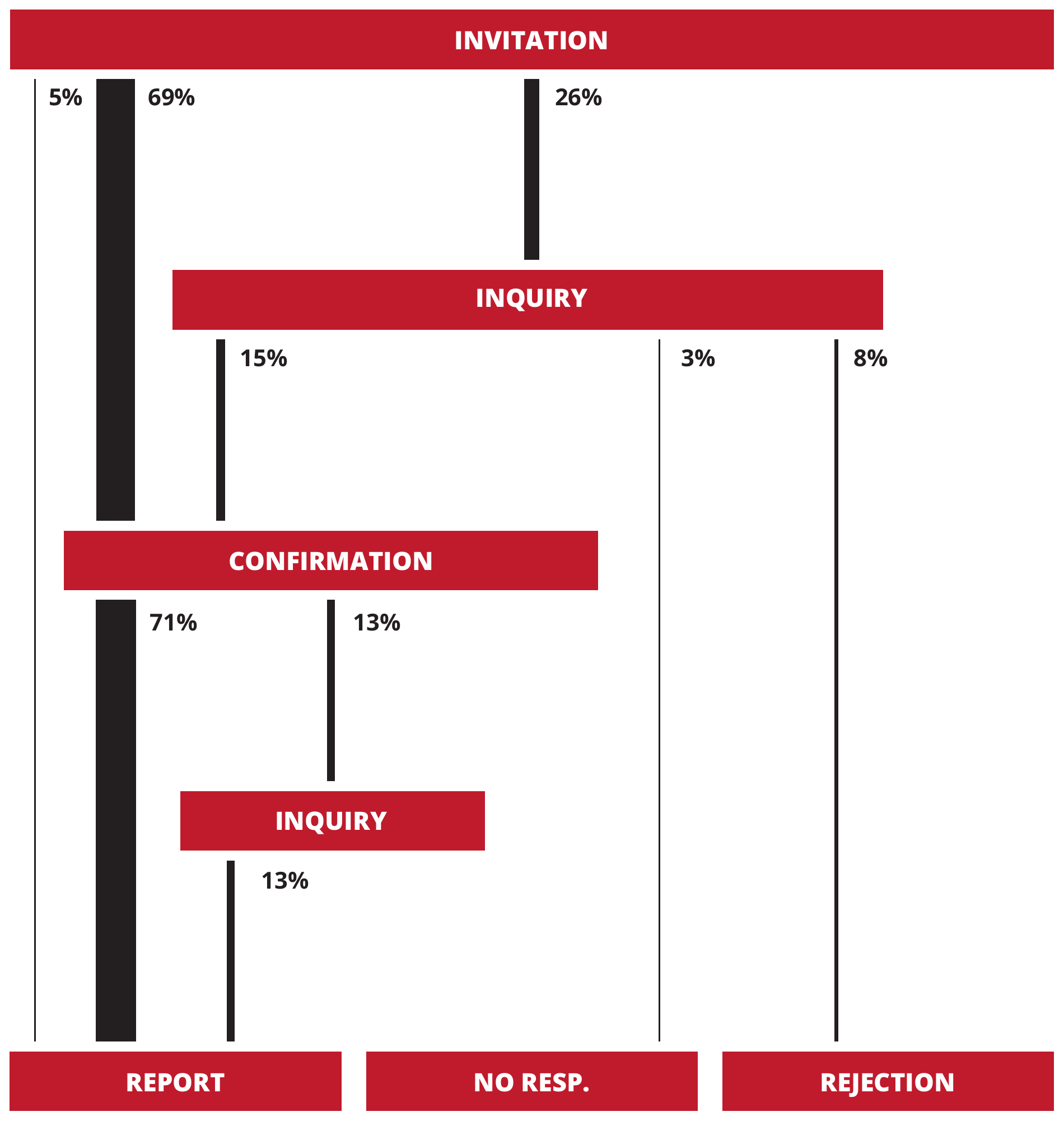}
	\caption{A graph corresponding to the review process with only \textbf{trusted} reviewers. Next to each edge are probabilities of a realisation of this process passing through the edge.}
	\label{fig:bdiag-trusted}
\end{figure}
A directed graph in which nodes correspond to phases and edges connect subsequent phases can be used as a visual representation of the review process. Graphs that describe our sample can be found in figures \ref{fig:bdiag-all}-\ref{fig:bdiag-trusted}. The value expressed in percent next to each edge is the probability that a realisation of the review process will pass through the edge - that is, the number of members from our sample for which the transition between nodes connected by the edge occurred divided by the size of the sample. Widths of edges were scaled proportionally to that probability.

What is striking is that only 43\% of all invitations actually result in a finished review (figure \ref{fig:bdiag-all}). Most of reviewers - that is 64\% - do not even respond to the initial invitation and 42\% ignore the inquiry. These poor results are mostly driven by reviewers that belong to the \textbf{other} group (figure \ref{fig:bdiag-other}), which constitutes the majority of all reviewers. Only 31\% of \textbf{other} reviewers finish the review, 73\% ignore the initial inquiry, 51\% do not answer at all and 16\% reject the invitation. On the other hand, \textbf{trusted} reviewers - who are in minority - are far more reliable. Most of them, 74\%, respond to the invitation and 89\% finish the review. Only 3\% do not answer and 8\% reject. As we will show in the following sections, this disparity between \textbf{trusted} and \textbf{other} reviewers may play a crucial role in the review process and is the key factor that determines its effectiveness.

\section{Review Times}

Review time, that is the number of days between the \textbf{invitation} phase and \textbf{report} phase, is the most direct and tangible measure of the efficiency of the review process. Since our sample contains information about the beginning and end of each phase, we were able to acquire distributions of review time for \textbf{trusted} and \textbf{other} reviewers, as well as partial distributions of days between all intermediate phases. These partial distributions are especially interesting, as they can serve as building blocks with which one can create a simulation of the entire review process and recreate the cumulative distribution of review time under various assumptions. 

\begin{figure}
	\centering
	\includegraphics[width=6cm]{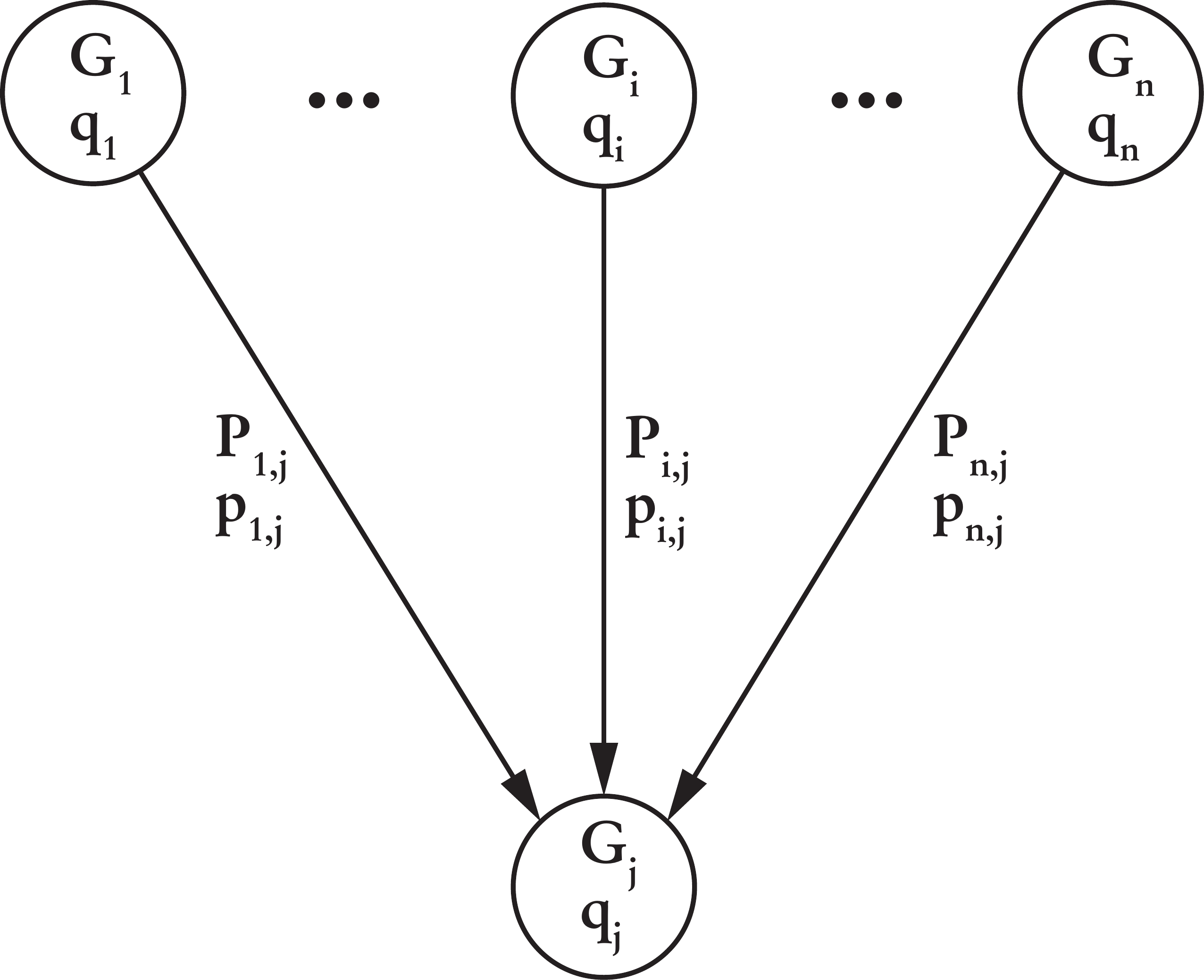}
	\caption{A schematic representation of a node from the review process graph, its predecessors and all associated probabilities.}
	\label{fig:bdiag-theory}
\end{figure}
The distribution of review time can be reassembled using partial distributions in the following way. To each node (phase) $j$ of the review process graph (figures \ref{fig:bdiag-all}-\ref{fig:bdiag-trusted}) one can assign the probability $q_j$ that a realisation of the process will pass through node $j$ and the probability distribution $G_j(t)$ of days between the \textbf{invitation} phase and phase $j$. Similarly, each edge is characterised by the probability $p_{i,j}$ that the review process will pass from phase $i$ to $j$ and the probability distribution $P_{i,j}(t)$ of days associated with such a transition. Given all these probabilities, $G_j(t)$ can be calculated as follows
\begin{equation}
	G_j(t) = \sum_{\{i\}_j} w_{i, j}\ (G_{i}\convolution P_{i, j})(t)\label{equ:gj}
\end{equation}
where the summation is over set $\{i\}_j$ of all predecessors of node $j$ and symbol $\convolution$ represents the discrete convolution
\begin{equation}
	(G_{i}\convolution P_{i, j})(t) = \sum_{t'=0}^{t} G_{i}(t') P_{i, j}(t-t').
\end{equation}
Weights $w_{i, j}$ are defined as
\begin{equation}
	w_{i, j} = \frac{q_i p_{i,j}}{q_j}.\label{equ:wij}
\end{equation}
and the probability $q_j$ can be expressed as
\begin{equation}
	q_j = \sum_{\{i\}_j} q_{i} p_{i, j}.\label{equ:qj}
\end{equation}
Equations \ref{equ:gj}-\ref{equ:qj} are recursive. The distribution $G_j(t)$ associated with node $j$ depends on the corresponding distributions associated with predecessors of node $j$ and probabilities $q_j$ exhibit similar dependence. As such, these equations can be solved recursively if one assumes appropriate initial conditions for nodes without parents (in our case it is $q_{\text{invitation}} = 1$ and $G_{\text{invitation}}(t)=\delta_{0,t}$ for the node that corresponds to the \textbf{invitation} phase) and acquires probabilities $P_{i, j}$ and $p_{i,j}$ from the sample. One last fact worth noting is that the quantity $q_i p_{i,j}$ from the numerator in equation \ref{equ:wij} is actually the same as the probability in figures \ref{fig:bdiag-all}-\ref{fig:bdiag-trusted} next to each edge.

\begin{figure}
	\centering
	\includegraphics[width=8cm]{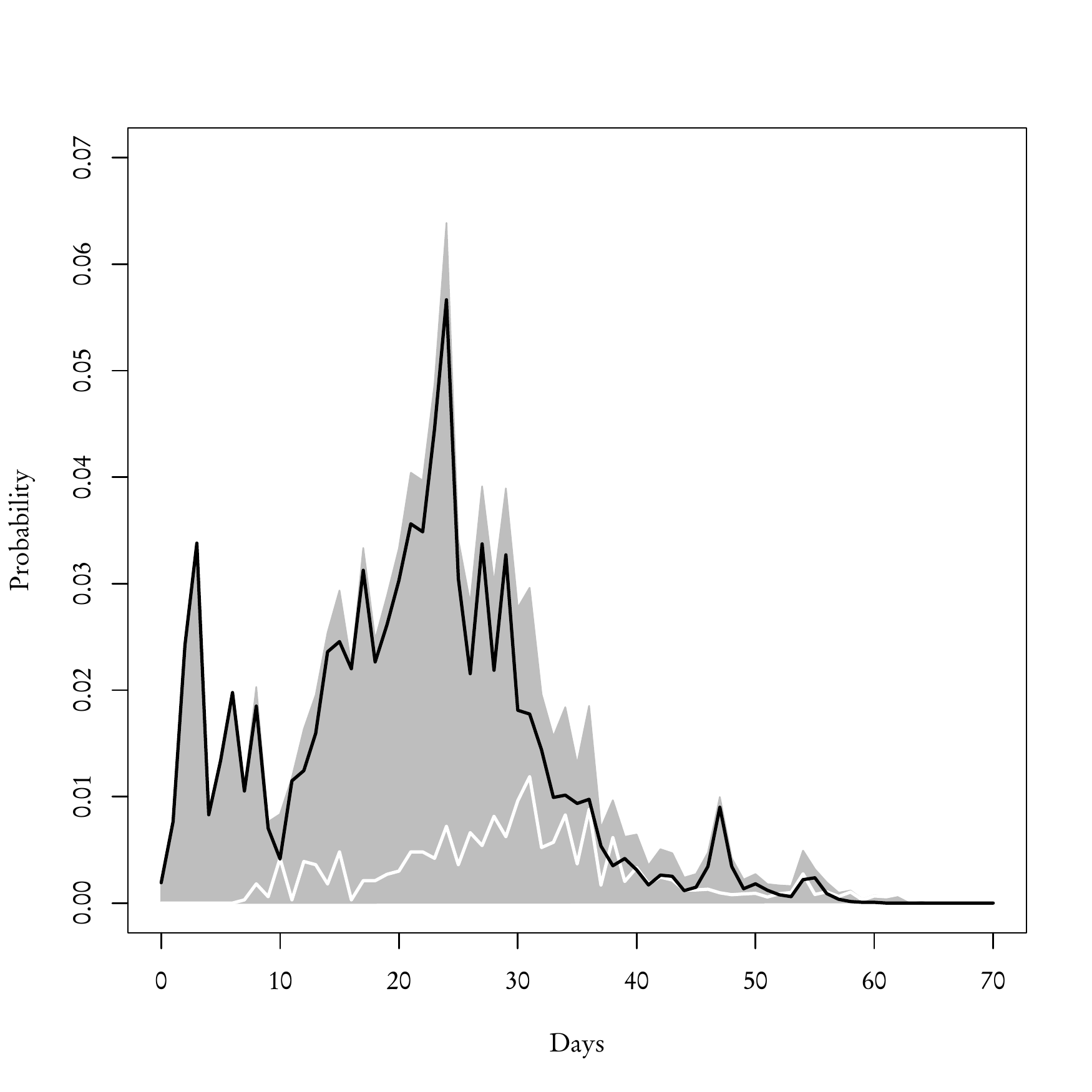}
	\caption{The theoretical probability distribution of review time for \textbf{trusted} reviewers who responded to the initial invitation (black line), who received an inquiry (white line) and their sum which gives the distribution for all \textbf{trusted} reviewers (filled polygon).}
	\label{fig:totdist-trusted-theor}
\end{figure}
\begin{figure}
	\centering
	\includegraphics[width=8cm]{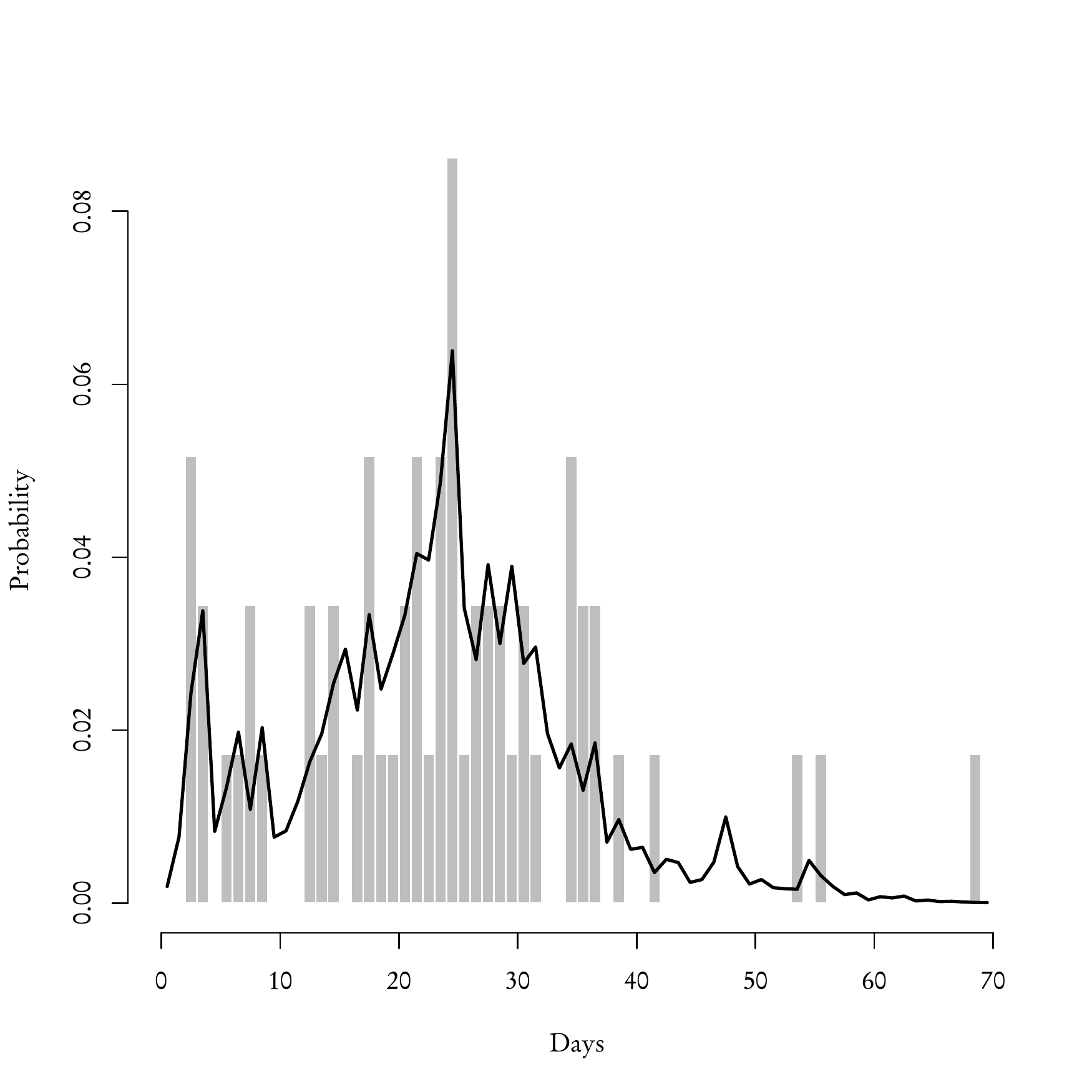}
	\caption{The probability distribution of review time for \textbf{trusted} reviewers: theoretical - black line, from data - grey bars.}
	\label{fig:totdist-trusted}
\end{figure}
\begin{figure}
	\centering
	\includegraphics[width=8cm]{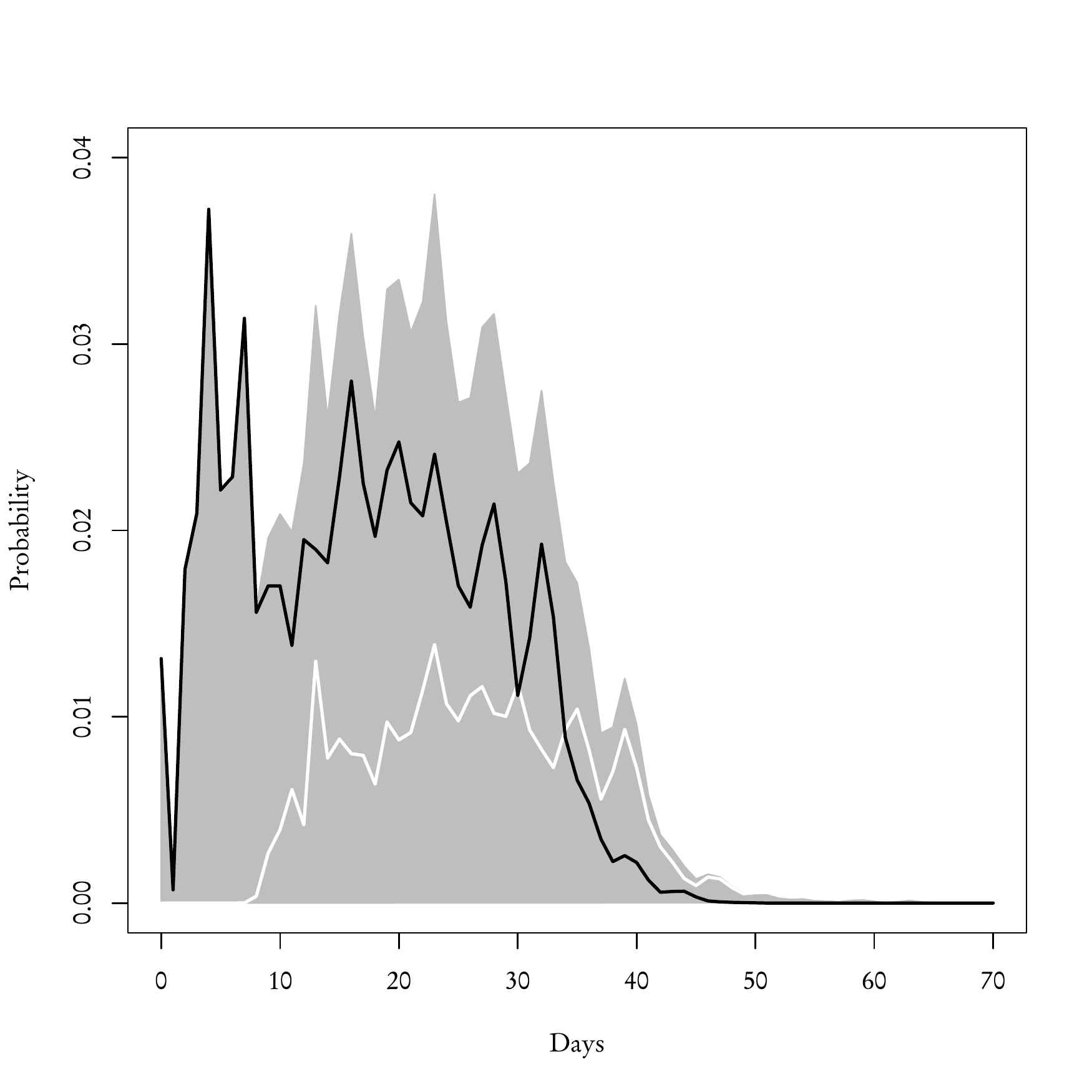}
	\caption{The theoretical probability distribution of review time for \textbf{other} reviewers who responded to the initial invitation (black line), who received an inquiry (white line) and their sum which gives the distribution for all \textbf{other} reviewers (filled polygon).}
\end{figure}
\begin{figure}
	\centering
	\includegraphics[width=8cm]{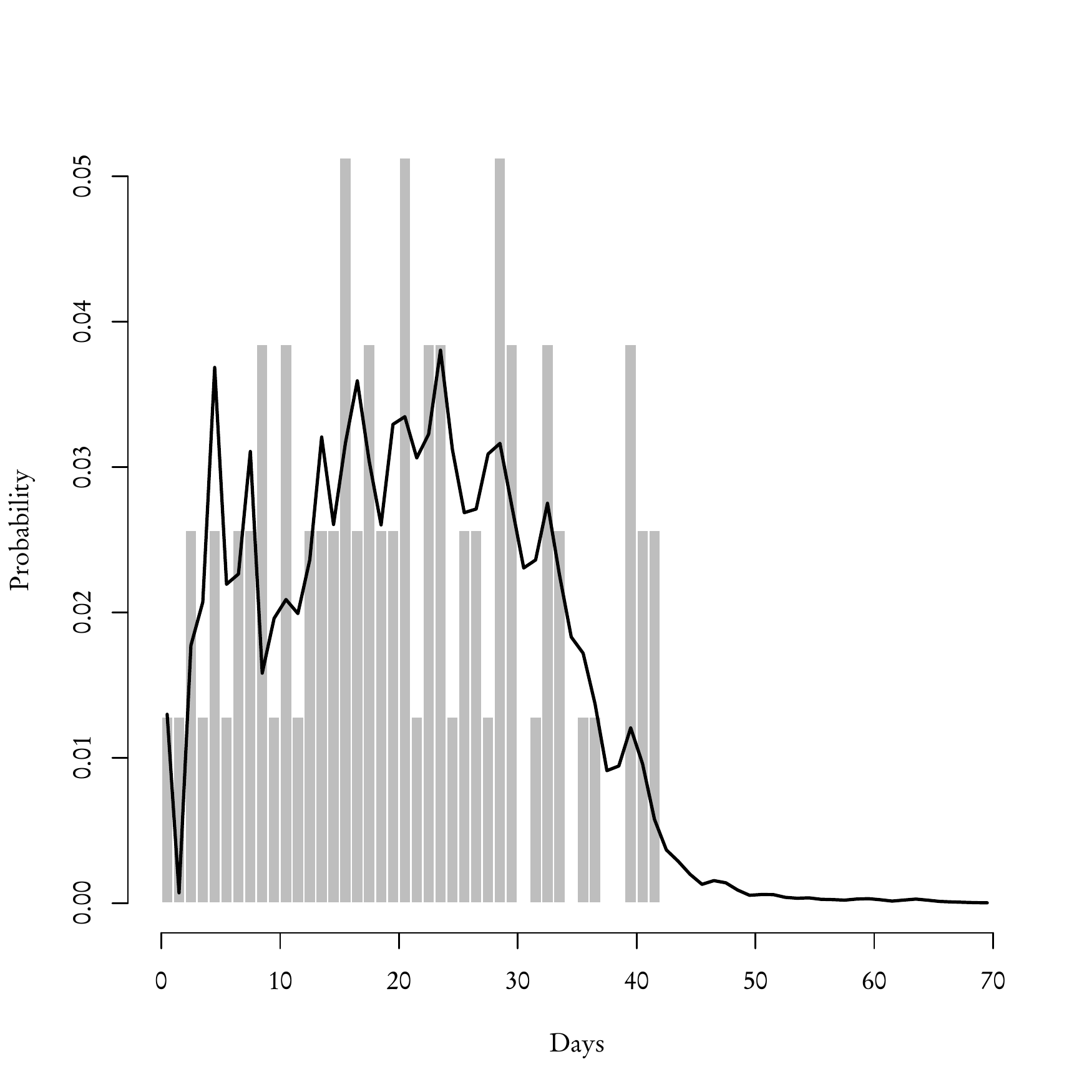}
	\caption{The probability distribution of review time for \textbf{other} reviewers: theoretical - black line, from data - grey bars.}
	\label{fig:totdist-other}
\end{figure}
Using the aforementioned procedure we recreated the distribution of review times for both \textbf{trusted} and  \textbf{other} reviewers which we then compared with the corresponding empirical distributions from the sample. According to our theoretical calculations based on equations \ref{equ:gj}-\ref{equ:qj} the average review time for \textbf{trusted} reviewers is 23 days with standard deviation of 12 days. Average review time and standard deviation acquired from the sample are the same. As for \textbf{other} reviewers, the theoretical average review time is 20 days with standard deviation of 11 days and the sample, again, yields the same values. One-sample Kolmogorov-Smirnov test performed to compare the theoretical distribution with the sample gives p-value 0.88 for \textbf{trusted} reviewers and 0.97 for \textbf{other} reviewers. It means that the distributions of review times calculated using partial distributions are essentially the same as the ones obtained directly from data. This is an important and non-obvious observation, as the only underlying assumption behind equations \ref{equ:gj}-\ref{equ:qj} is that the review process is memoryless - that is partial distributions assigned to edges do not depend on the history of the process. Results presented thus far seem to confirm this assumption and it is reinforced even further in the following section.

Other than the validity of theoretical distributions, there are two main conclusions that can be drawn from results presented in figures \ref{fig:totdist-trusted-theor}-\ref{fig:totdist-other}. Firstly, the review time distribution is bimodal. Reviewers who either confirmed or sent in their reviews after receiving the invitation are the ones who contribute to the first maximum (and they are in the majority of those who actually completed the reports - 69\% of \textbf{other} and 82\% of \textbf{trusted}). Secondly, distributions of review time are similar for \textbf{trusted} and \textbf{other} reviewers. The difference between means and standard deviations is negligible from any practical standpoint and two-sample Kolmogorov-Smirnov test for both empirical distributions gives p-value 0.40. Based on these fact one can make a very strong assumption that the distribution of review time is the same across the entire population of reviewers and does not depend on the type of reviewer.

\section{Simulations of the review process}

So far we have considered review times of a single reviewer. However, editors usually need more than one review in order to judge whether to publish an article. In the case of our data from JCSC, the editor required two reviews per article and sent invitations to five reviewers on average - one \textbf{trusted} and four \textbf{other}. While this review strategy indeed resulted in two reviews per article on average (2.34 to be exact), 9 articles were published after receiving only one review, 24 after 2 reviews, 21 after 3 and 4 after 4 reviews. This discrepancy between the target number of reviews and the number of reviews actually received stems from the difference in the probability of finishing the report between \textbf{trusted} and \textbf{other} reviewers. We are going to call this probability the completion rate.

\begin{figure}
	\centering
	\includegraphics[width=8cm]{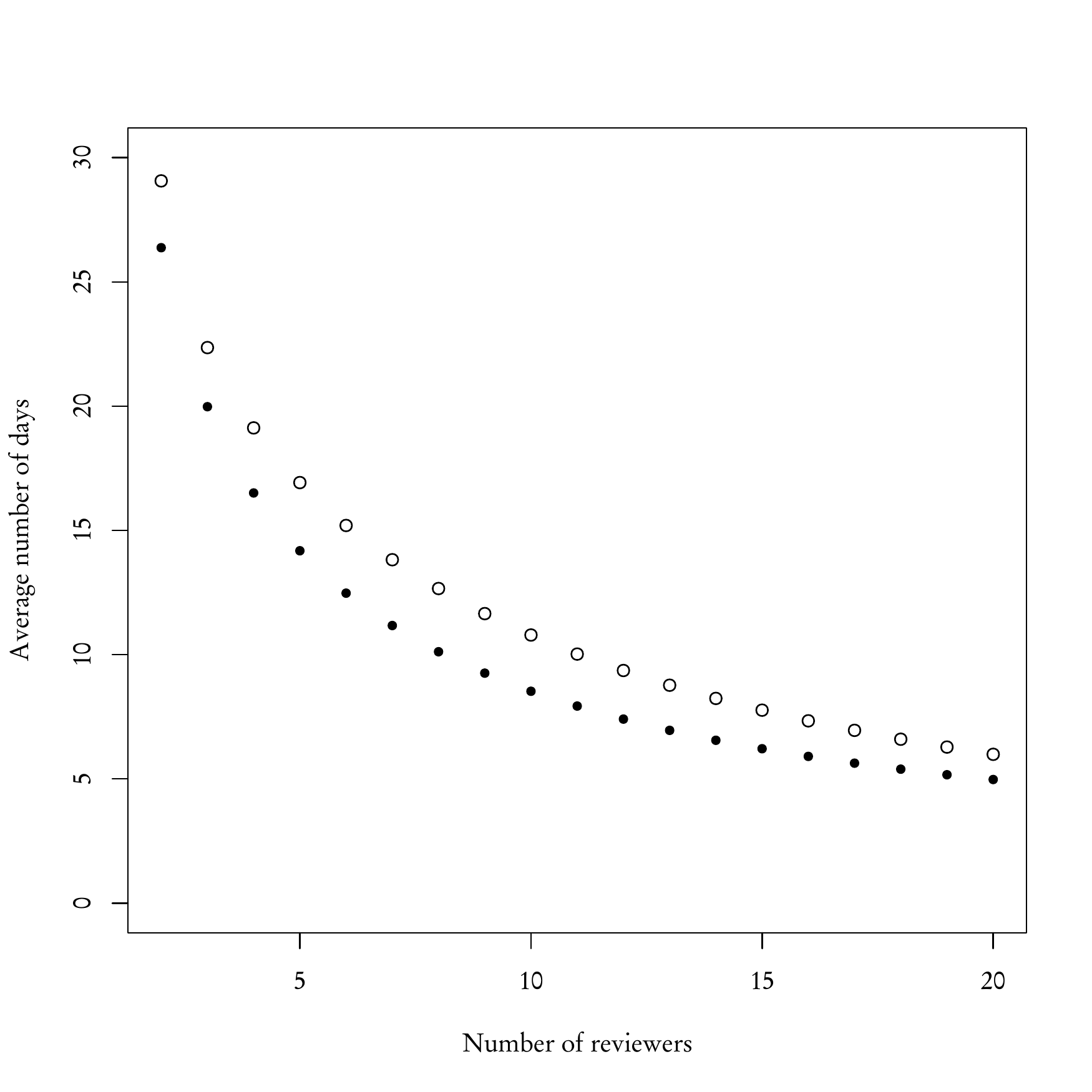}
	\caption{Average time of acquiring two reviews for \textbf{trusted} (empty circles) and \textbf{other} (filled black circles) reviewers when all reviewer finish their reviews.}
	\label{fig:allways-finish}
\end{figure}
Using partial distributions we can easily simulate the effects of any editorial strategy and find the number of reviewers needed to achieve a certain number of reviews per article. We will use the average time of receiving two reviews as a measure of effectiveness of each strategy. Figure \ref{fig:allways-finish} shows these average times under the assumption that a reviewer always writes the report (the completion rate is equal 1) for both \textbf{trusted} and \textbf{other} reviewers as a function of the number of reviewers. The average time decreases as the number of reviewers increases and results for \textbf{trusted} and \textbf{other} reviewers are very similar. This is intuitive and consistent with our prediction made in the previous section.

\begin{figure}
	\centering
	\includegraphics[width=8cm]{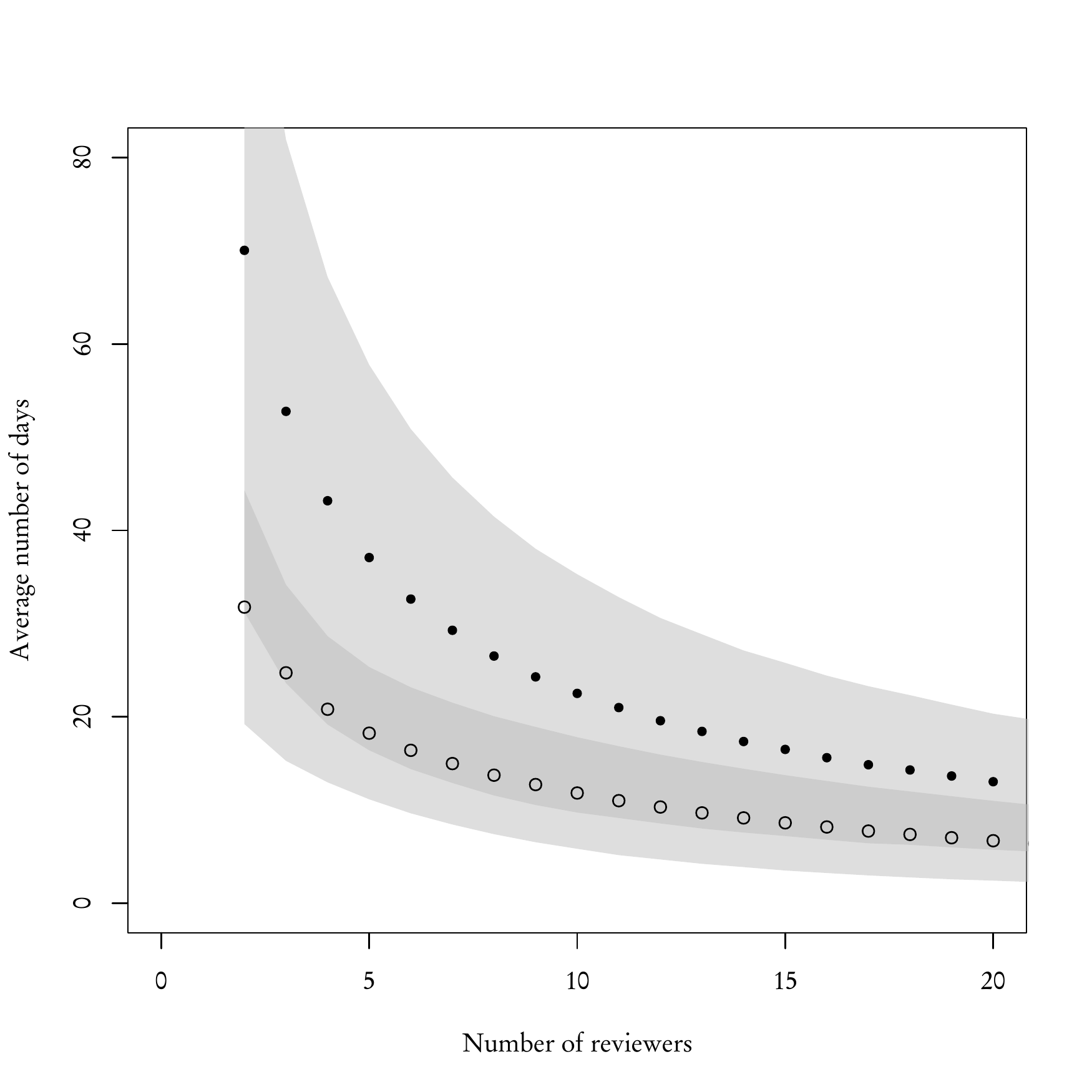}
	\caption{Average time of acquiring two reviews for \textbf{trusted} (empty circles) and \textbf{other} (filled black circles) reviewers with completion rate taken into account. Filled polygon represents standard deviation.}
	\label{fig:fraction-finish}
\end{figure}
\begin{figure}
	\centering
	\includegraphics[width=8cm]{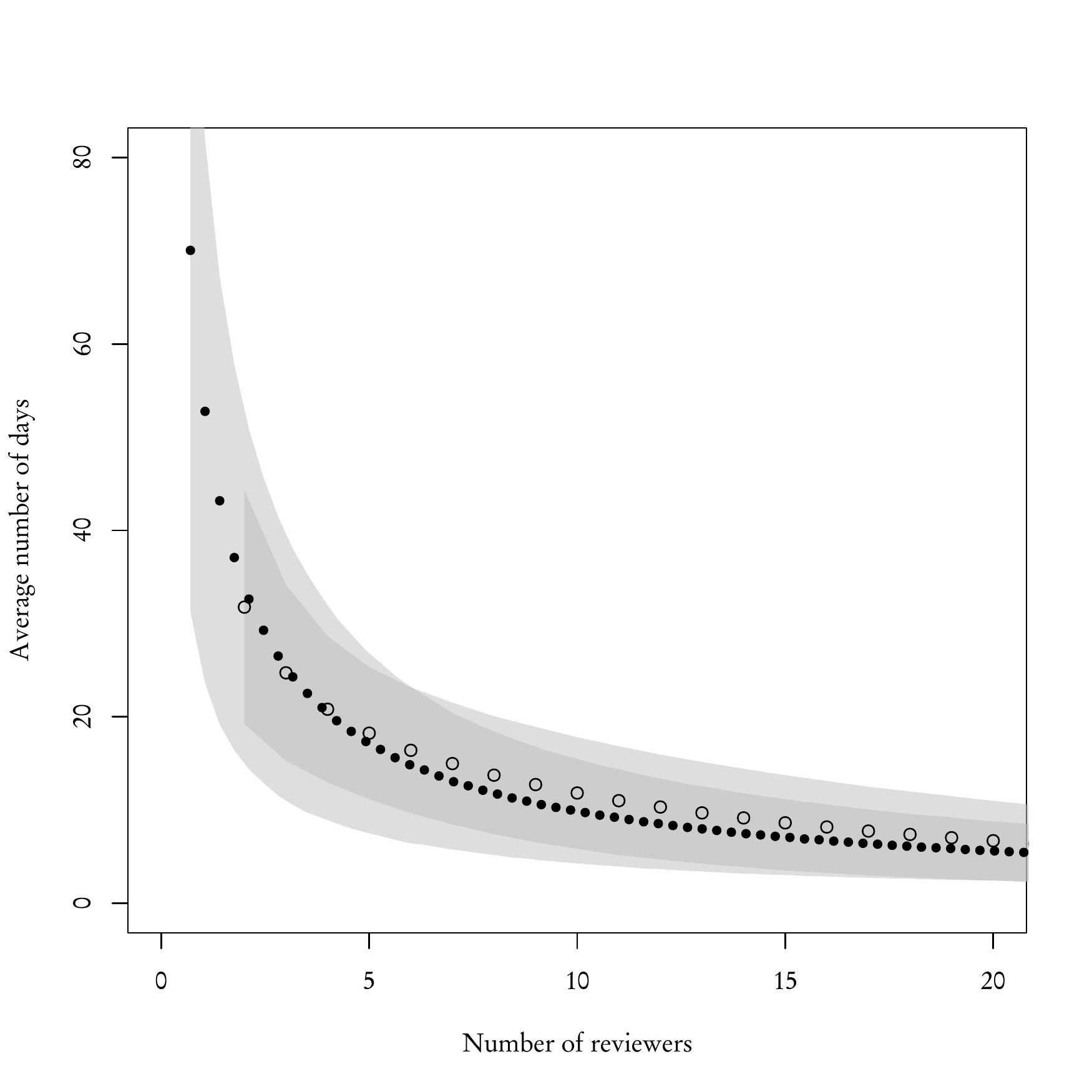}
	\caption{Same as figure \ref{fig:fraction-finish} but with the X axis rescaled for \textbf{other} reviewers.}
	\label{fig:fraction-finish-rescaled}
\end{figure}
The assumption that reviewers always write the report is not realistic. If we want to take into account the fact that the actual completion rate of the review process for a single reviewer is much smaller, especially for \textbf{other} reviewers, then some additional strategy needs to be introduced to deal with situations when two reviews are not received at all. In our simulations we decided to use a very simple solution - if two reviews are not received, then invitations are resent to the same number of reviewers. This procedure is repeated if necessary until reviewers produce two reports in total. While this is not the most effective and time-efficient strategy, it still allows us to study the consequences of the difference between the completion rates of \textbf{trusted} and \textbf{other} reviewers.

Figure \ref{fig:fraction-finish} is analogous to figure \ref{fig:allways-finish} - in that it shows the average time of receiving two reviews - but this time we used the actual completion rates taken from the sample (89\% for \textbf{trusted}, 31\% for \textbf{other} reviewers) and employed the policy described in the previous paragraph. As can be clearly seen, the difference in completion rates between \textbf{trusted} and \textbf{other} reviewers results in a completely  different dynamics. \textbf{Other} reviewers are far less effective and their average times are much higher - for example, two reviews can be received from 2 \textbf{trusted} reviewers after 32 days but 2 \textbf{other} reviewers finish the reviews after 70 days. Even as the number of reviewers increases, this difference remains significant.

However, in the last section we have shown that distributions of review time for \textbf{trusted} and \textbf{other} reviewers are very similar which suggests that the completion rate is the leading factor during the review process. This claim is partially supported by results presented in figure \ref{fig:allways-finish}. If that is indeed the case, then one \textbf{trusted} reviewer should be ''worth'' 89\%/31\% \textbf{other} reviewers and conversely one \textbf{other} reviewer is ''worth'' 31\%/89\% \textbf{trusted} reviewers. By ''worth'' we mean that proportionally substituting one type of reviewer for another should yield the same results. Figure \ref{fig:fraction-finish-rescaled}, where the X axis for one type of reviewers was rescaled to match their worth in the other type of reviewers, confirms this prediction. The average number of days after which 2 reviews are acquired are similar and standard deviations, while not exactly the same - which is to be expected - are comparable.

\begin{figure}
	\centering
	\includegraphics[width=8cm]{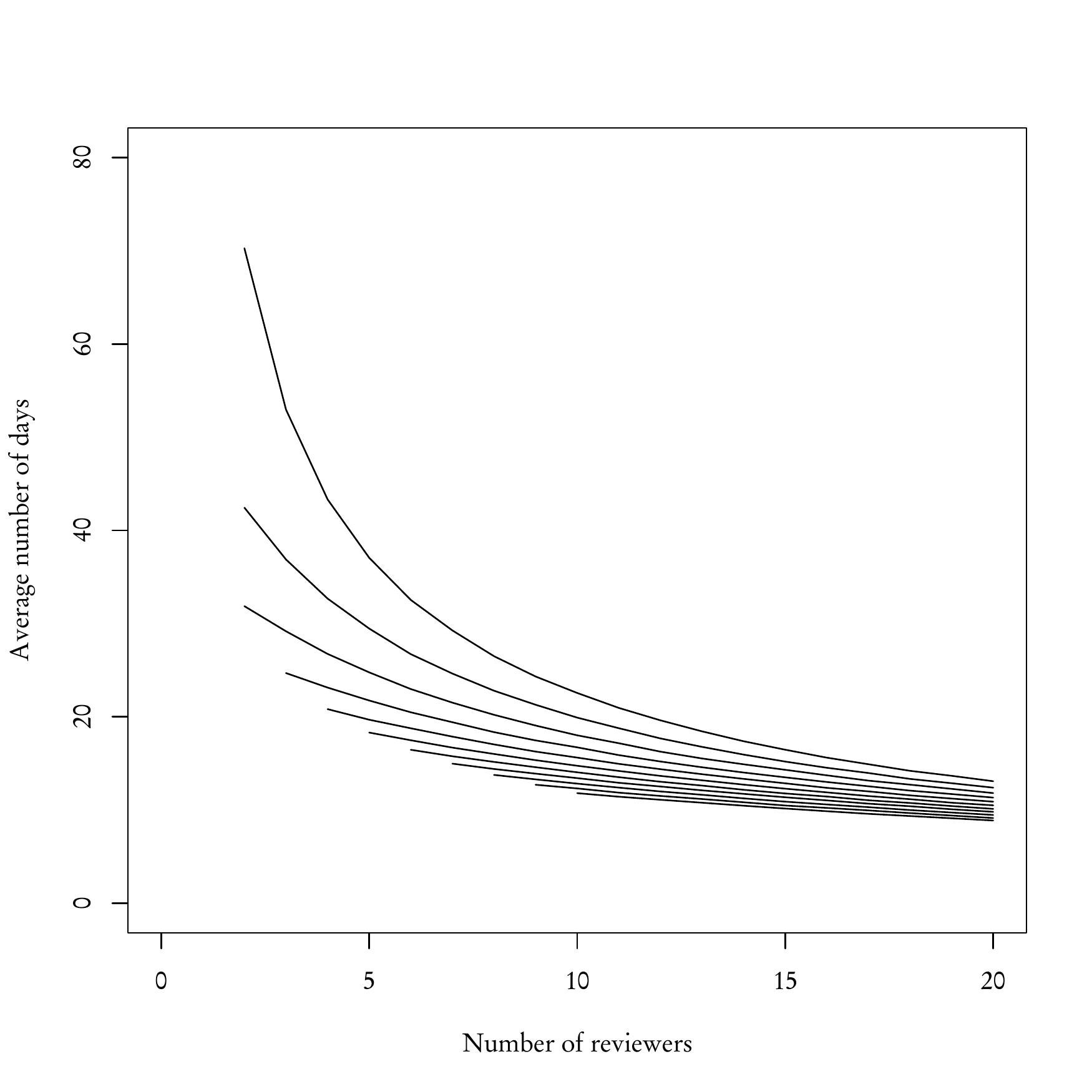}
	\caption{Average time of acquiring two reviews for a group of mixed reviewers. The X axis - total number of reviewers. Curves correspond to various numbers of \textbf{trusted} reviewers: 0 \textbf{trusted} - top curve, 10 \textbf{trusted} - bottom curve.}
	\label{fig:mix}
\end{figure}

\begin{figure}
	\centering
	\includegraphics[width=8cm]{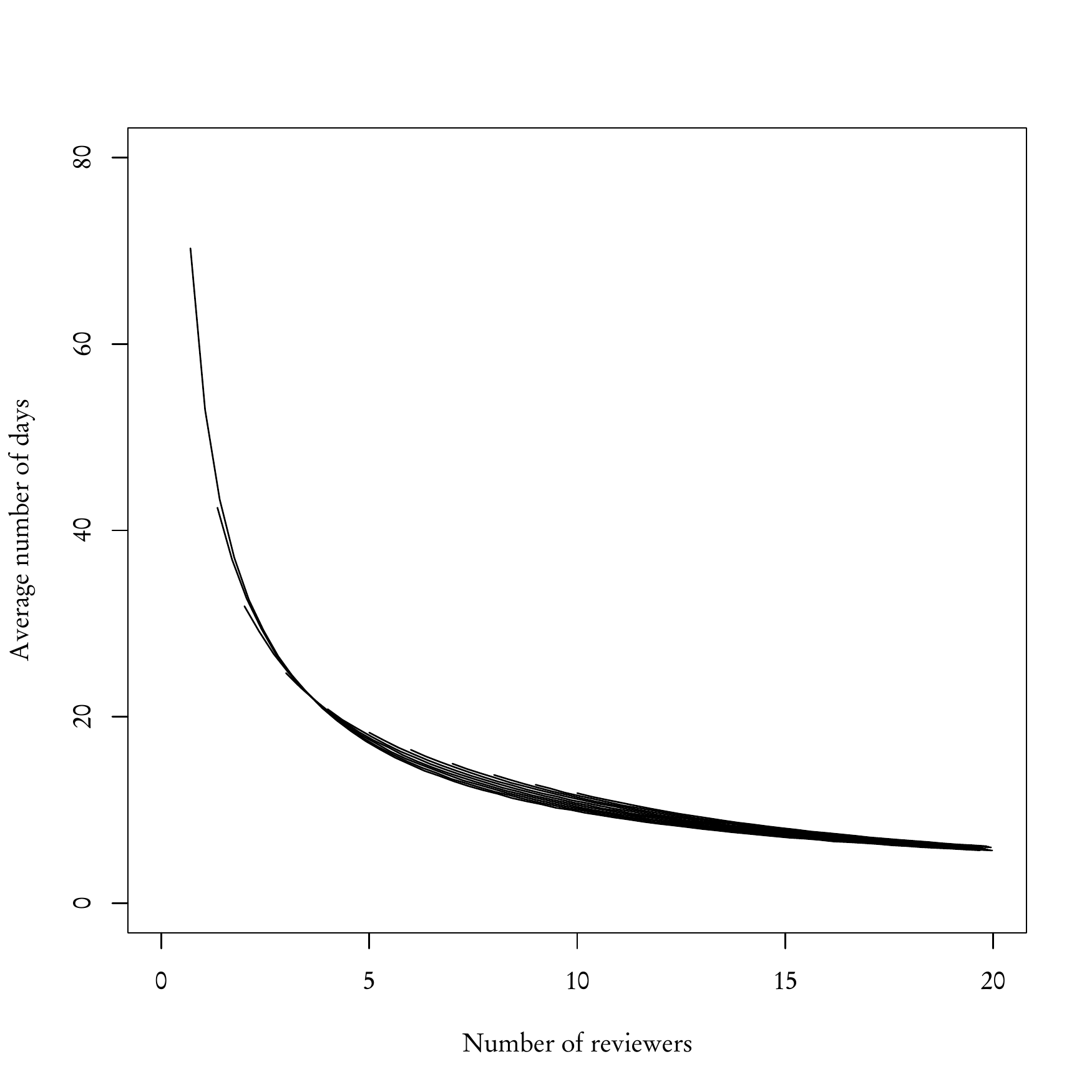}
	\caption{Same as figure \ref{fig:mix} but with rescaled X axis.}
		\label{fig:mix-rescaled}
\end{figure}

\begin{table}[t]
\centering
\begin{tabular}{r||rrrrrrrrrr}
  \hline
 & 0 & 1 & 2 & 3 & 4 & 5 & 6 & 7 & 8 & 9 \\ 
  \hline\hline
0 &   X &   X &  32 &  25 &  21 &  18 &  16 &  15 &  14 &  13 \\ 
  1 &   X &  42 &  29 &  23 &  20 &  17 &  16 &  14 &  13 &  12 \\ 
  2 &  70 &  37 &  27 &  22 &  19 &  17 &  15 &  14 &  13 &  12 \\ 
  3 &  53 &  33 &  25 &  20 &  18 &  16 &  15 &  13 &  12 &  11 \\ 
  4 &  43 &  29 &  23 &  19 &  17 &  15 &  14 &  13 &  12 &  11 \\ 
  5 &  37 &  27 &  22 &  18 &  16 &  15 &  14 &  13 &  12 &  11 \\ 
  6 &  33 &  25 &  20 &  17 &  16 &  14 &  13 &  12 &  11 &  10 \\ 
  7 &  29 &  23 &  19 &  17 &  15 &  14 &  13 &  12 &  11 &  10 \\ 
  8 &  26 &  21 &  18 &  16 &  14 &  13 &  12 &  11 &  11 &  10 \\ 
  9 &  24 &  20 &  17 &  15 &  14 &  13 &  12 &  11 &  10 &  10 \\ 
   \hline
\end{tabular}
\caption{Average number of days needed to receive two reviews from a group of reviewers with a given number of \textbf{trusted} (columns) and \textbf{other} (rows) reviewers. Values for groups of reviewers smaller than two were omitted.}
\label{fig:mix-table}
\end{table}
So far we have studied separately \textbf{trusted} and \textbf{other} reviewers, however the group of reviewers invited to review an article usually contains reviewers of both kinds. Figure \ref{fig:mix} shows the average time of acquiring two reviews when reviewer types are mixed. As one could expect, the average time decreases with the increasing total number of reviewers and \textbf{trusted} reviewers are far more effective than \textbf{other}. Still, by rescaling the X axis - that is by expressing the worth of one kind of reviewer using another - we get similar results (figure \ref{fig:mix-rescaled}). 

\begin{figure}
	\centering
	\includegraphics[width=9cm]{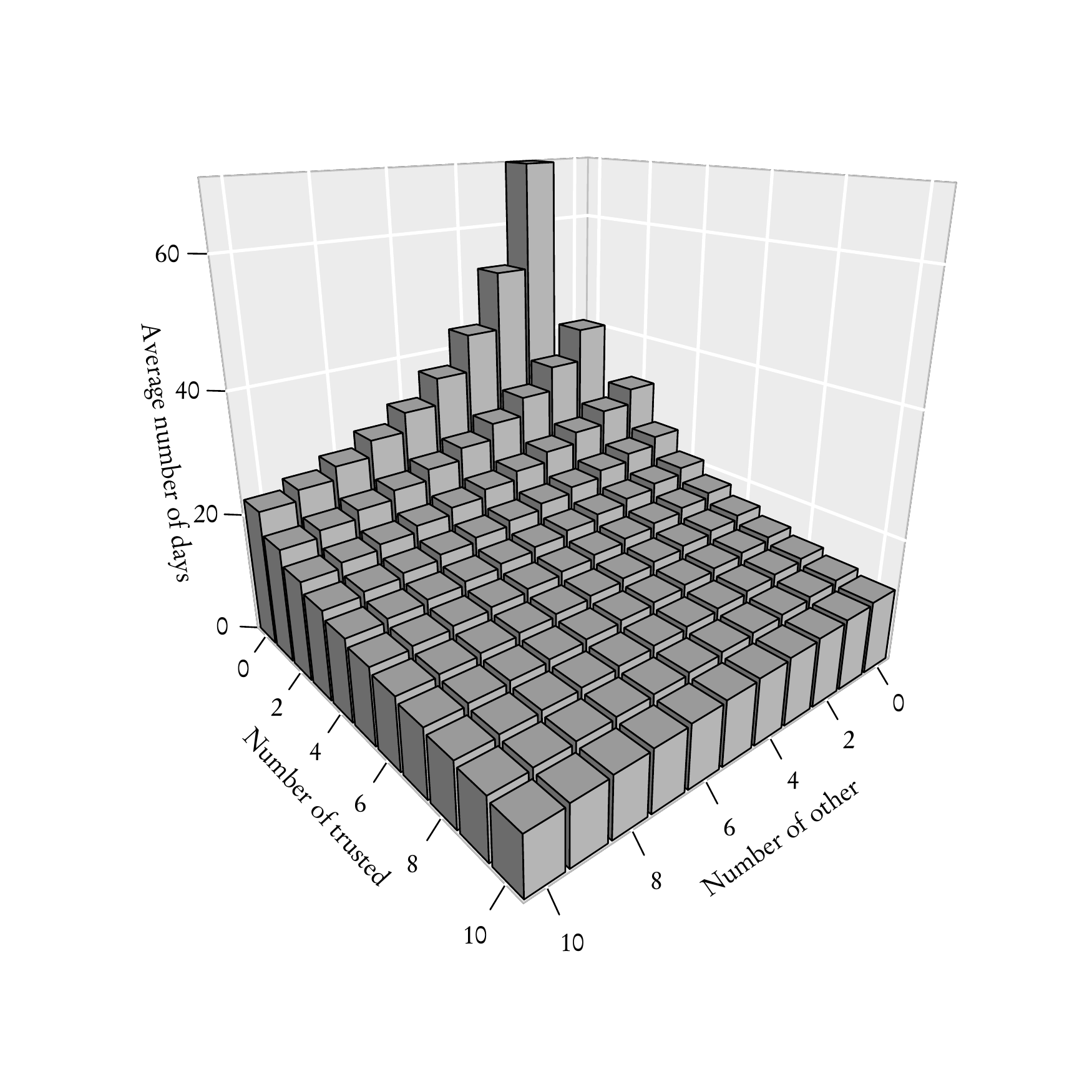}
	\caption{Average time of acquiring two reviews for a group of mixed reviewers.}
	\label{fig:mix-3d}
\end{figure}

	Information about average times in groups of mixed reviewers, expressed in a slightly different way in figure \ref{fig:mix-3d} and summarised in table \ref{fig:mix-table} can potentially be of great importance for editors and act as a guide in determining the optimal number of reviewers. For example, in order to receive two reviews after about 30 days, one needs to invite 7 \textbf{other} reviewers, 2 \textbf{trusted} or a mixed group of 4 \textbf{other} and 1 \textbf{trusted}. That last option is consistent with the choice made by the editor of JCSC who provided us with the data.

It is important to note that while editors may be tempted to invite only \textbf{trusted} reviewers - which would lead to shortest review times - such a policy would not only be not realistic but also inadvisable. Since the pool of potential \textbf{trusted} reviewers is limited, editors would be forced to invite the same reviewers multiple times within a short time frame. This, in turn, could discourage reviewers and make them more likely to turn down invitations.
	

\section{Discussion}

Our results show that the distribution of review time is similar for all kinds of reviewers and it is the completion rate that is the main factor that determines the effectiveness of the review process. \textbf{Trusted} reviewers, that is reviewers known personally by the editor, are far more reliable than \textbf{other} reviewers. Their completion rate is very high, which means that they are much more likely to answer the invitation and finish the review. On the other hand, only a fraction of \textbf{other} reviewers answer the initial invitation and write the report. It means that \textbf{trusted} reviewers are objectively better than \textbf{other} reviewers and there is no advantage in choosing the latter over the former. In an ideal world, editors would invite only \textbf{trusted} reviewers, which, unfortunately, is not possible. 

One question remains, then - who exactly is this mythical \textbf{trusted} reviewer? What makes the difference between \textbf{trusted} and \textbf{other} reviewers? In the case of JCSC, it was a personal relationship with the editor. One can easily imagine that this mechanism works in a very similar way in journals of comparable scope. What about bigger journals or ones in which editors do not choose reviewers themselves? Even without knowing the editor, reviewers invited by prestigious journals with high impact factor may be more inclined to write the review and thus act as \textbf{trusted}. In the end, it seems that the distinction between \textbf{trusted} and \textbf {other} reviewers is slightly artificial and was motivated mostly by the way our data is structured. Instead, the completion rate is a much more intrinsic property that differentiates between reviewers. It is also important to notice that the completion rate is not a property of a reviewer, but of his relationship with other entities - be it journals, editors or even other reviewers. As such, the same reviewer can be treated as \textbf{trusted} by some journals and as \textbf{other} by others. Also, since relations between people can change, the completion rate does not have to be constant and it may evolve with time.

Authors of manuscripts, reviewers and editors form a complex network of mutual connections, the structure of which have a direct influence on the effectiveness of the review process. However, since editors are the ones who actually manage the entire process, it would seem that their workflow is equally, if not even more important. With the right kind of workflow one can potentially overcome many shortcoming of the behaviour of both authors and reviewers.	We have shown that through very naive and most certainly not optimal means - by sending invitations to a certain number of potential reviewers - it is possible to achieve almost any desirable average review time. While it is a very simple example, our results presented in this manuscript can be used as a foundation necessary to study the dynamics of the review process and determine the optimal workflow for an editor, which will be the subject of our future work.

\section{Acknowledgements}

A.F. \& P.F. were supported by the Foundation for Polish Science (grant no. POMOST/2012-5/5) and by the European Union within European Regional Development Fund (Innovative Economy). This paper is a part of scientific activities in COST Action TD1306 New Frontiers of Peer Review (PEERE).

%Each link in the graph represents a single stage of the process and its two-dimensional weight reflects duration  and the probability of occurrence of the stage.  
%In this sample, the names of the referees have been anonymized (replaced by the numbers).
%The result of these combined factors is that reviewers are often overloaded and frustrated with the increasing demands of the peer review system.

%Perhaps one of the reasons why reviewers are overburdened is because editors call on them more frequently than they probably should. Given that there is a finite
%amount of reviewer time available, we need a system that capitalizes effectively on�but does not waste�reviewers� time and energy. 


\begin{thebibliography}{99}
\bibitem{Wager_2001} Wager E., Jefferson T., {\it The shortcomings of peer review}, Learned Publishing, {\bf 14}, pp.257-263 (2001).
\bibitem{Cooper_2009} Cooper M.L., {\it Problems, pitfalls, and promise of the peer-review process: Commentary on Trafimow \& Rice (2009)}, Perspect. Psychol. Sci. {\bf 4}, 84�90 (2009).
\bibitem{Baker_2002} Baker D., {\it The peer review process in science education journals}, Research in Science Education, {\bf 32}, 171�180 (2002).
\bibitem{PRC_2008} Publishing Research Consortium, {\it Peer review in scholarly journals: Perspective of the scholarly community � an international study}, (2008).
\bibitem{Bornmann_2011} Bornmann L., {\it Scientific peer review}, Ann. Rev. Inf. Sci. Technol. {\bf 45}, pp.199-245 (2011).
\bibitem{Cawley_2011} Cawley V., {\it An Analysis of the Ethics of Peer Review and Other Traditional Academic Publishing Practices}, Int. J. Soc. Sci. Human., {\bf 1}, 205-213 (2011).
\bibitem{Resnik_2008} Resnik D.B, Gutierrez-Ford C., Peddada S., {\it Perceptions of ethical problems with scientific journal peer review: an exploratory study}, Sci. Eng. Ethics. {\bf 14}, 305-310 (2008).
\bibitem{Schwartz_2009} Schwartz S.J., Zamboanga B.L., {\it The Peer-Review and Editorial System: Ways to Fix Something That Might Be Broken}, Perspect. Psychol. Sci., {\bf 4}, 54-61 (2009).
\bibitem{Kravitz_2010} Kravitz R.L., Franks P., Feldman M.D., et al., {\it Editorial Peer Reviewers' Recommendations at a General Medical Journal: Are They Reliable and Do Editors Care?}, PLoS ONE , {\bf 5}, e10072 (2010).
\bibitem{Newton_2010} Newton D.P., {\it Quality and Peer Review of Research: An Adjudicating Role for Editors}, Account. Res., {\bf 17}, 130�145 (2010).
\bibitem{COPE} Committee on Publication Ethics, {\it COPE Ethical Guidelines for Peer Reviewers}, \url{http://publicationethics.org/files/Ethical_guidelines_for_peer_reviewers_0.pdf}, (2013). 
\bibitem{Wager_2006} Wager E., {\it Ethics: What is it for?}, Nature: Web Debate � Peer-review, \url{http://www.nature.com/nature/peerreview/debate/nature04990.html}, (2006).


\end{thebibliography}
\end{document}